\documentclass[sigconf]{acmart}

\usepackage{booktabs} % For formal tables

\usepackage[]{algorithm2e}
\usepackage{array}
\usepackage{dblfloatfix}    % To enable figures at the bottom of page
\usepackage[tight,footnotesize]{subfigure}
\usepackage{caption}
\usepackage{url}
\usepackage{paracol}
\usepackage{ragged2e}
\usepackage{balance}

\setcopyright{acmlicensed}
\acmDOI{10.475/123_4}

\acmISBN{123-4567-24-567/08/06}

\acmConference[FSE'17]{ACM FSE conference}{September 2017}{Paderborn, Germany}
\acmYear{2017}
\copyrightyear{2017}
\acmPrice{15.00}
\acmDOI{10.1145/3106237.3122818}
\acmISBN{978-1-4503-5105-8/17/09}
\acmConference[ESEC/FSE'17]{2017 11th Joint Meeting of the European Software Engineering Conference and the ACM SIGSOFT Symposium on the Foundations of Software Engineering}{September 4--8, 2017}{Paderborn, Germany} 
\acmYear{2017}
\copyrightyear{2017}
\acmPrice{15.00}
\begin{document}

\title{Cherry-Picking of Code Commits in Long-Running, Multi-release Software}
\author{Panuchart Bunyakiati and Chadarat Phipathananunth}
\affiliation{%
  \institution{University of the Thai Chamber of Commerce}
}
\email{panuchart_bun, chadarat_phi@utcc.ac.th}

\begin{abstract}
This paper presents Tartarian, a tool that supports maintenance of software with long-running, multi-release branches in distributed version control systems. When new maintenance code, such as bug fixes and code improvement, is committed into a branch, it is likely that such code can be applied or reused with some other branches. To do so, a developer may manually identify a commit and cherry pick it. Tartarian can support this activity by providing commit hashtags, which the developer uses as metadata to specify their intentions when committing the code. With these tags, Tartarian uses dependency graph, that represents the dependency constraints of the branches, and Branch Identifier, which matches the commit hashtags with the dependency graph, to identify the applicable branches for the commits. Using Tartarian, developers may be able to maintain software with multiple releases more efficiently.
\end{abstract}



\maketitle
\section{Introduction}

Distributed, branch-based version control systems, such as Github, have increasingly gained popularity in the software community \cite{Kalliamvakou15}. The release-branching feature in Github allows software to be maintained as multiple releases. It is observed that projects such as Jetty and Python maintain multi-release branches because of the dependency constraints on external libraries. When new maintenance code is committed into one release branch, it is likely that such code can be reused with some other branches. Tartarian can identify the branches that can benefit from reusing the code and may help developers to efficiently maintain software. 

\section{Related works}

Code maintenance, especially bug fixing, is studied \cite{Aranda09} and the results suggest that bug management should not consider the stage of bug (active, inactive, etc.) and workflow but should be based on satisfying \emph{``stakeholder's goals during the lifespan of bug''} such as assignment of ownership and search for knowledges. Guo et al. \cite{Guo10} study bugs in Windows Vista and Windows 7 and develop a statistical model to predict the probability of a bug to be fixed; and similarly, bug reporting and fixing are found to be related to individuals, social and organisational issues. 

Several works (e.g. \cite{Casey13} and \cite{Cortes-Coy14}) study the automatic generation of commit messages. Jiang and McMillan \cite{Jiang17} propose a tool that automatically generate commit messages in the \emph{``verb+object"} format by classifying a diff on the commit messages. Instead of creating the commit message automatically, our work provides a set of hashtags for developers to annotate the commit messages, which may enhance code reuse through better semi-automated communication. Our approach focuses on team collaboration and is different from code clone detection techniques \cite{Bellon} and \cite{Roy} that analyze code to identify similarities between branches to support code reuse. 

\section{Cherry-Picking}

Cherry-picking is a process to manually pick commits in one branch and apply them into another branch. The difficulty lies in the fact that developers must know which commits and which branches the commits should be applied. Given the condition that those branches cannot be merged as they are long-running releases, the developers can use the function \emph{git cherry-pick} to reuse the code and apply the commits into their branches.

To illustrate the concept, the direct acyclic graph (DAG) of a repository below has three releases 1.1, 1.2 and 1.3 where release 1.1 has commit a, release 1.2 has commits a, b and c and release 1.3 has commits a, b, c, d and e. Once the bug fix f is committed to release 1.3 (with the hash \emph{db55fd2}), it can be cherry picked to release 1.1 and 1.2 hence release 1.1 becomes a and f, release 1.2 becomes a, b, c and f and release 1.3 becomes a, b, c, d, e and f.

\scriptsize
\begin{verbatim}
* b44ea88 (release1.2) Added f (cherry picked from db55fd2)
| * 4fcd797 (release1.1) Added f (cherry picked from db55fd2)
| | * db55fd2 (HEAD -> master, release1.3) Added f
| | * 3e63476 Added e
| | * f0a729e Added d
| |/
|/|
* | 9981679 Added c
* | 01ea7e6 Added b
|/
* 5597940 Added a
\end{verbatim}
\normalsize

\section{Current Practices}

Regardless of the workflow that the developers are using, we argue that once there is a necessary for cherry-picking, there requires a tool that help to automate the decision making by identifying affected release branches. We aim to develop Tartarian to apply to any kind of branch-based workflow that allows cherry-picking. This section describes existing practices related to cherry-picking.

First, to consider which commit to be cherry picked, developers can use \emph{git cherry} to find commits to be applied from the \emph{topic} branch into the \emph{upstream} branch, reporting as \emph{+} or \emph{-}. This approach only considers code diff and does not consider dependency constraints; hence, all commits with code diff are simply reported even though they are indeed inapplicable to the branch. 

Second, to keep track of the cherry-pick, developers must use the option to record the original commit using \emph{-x}; with this option, Git appends the commit message with \emph{``cherry picked from commit \textless commitid\textgreater''}. This helps to keep track of the cherry-pick so that  other developers can use the command \emph{git branch $--$contains \textless commitid\textgreater} to check which of the branches have the commit and use the command \emph{git log $--$ \textless file path\textgreater} to track the changes to the file. If the commit is done without \emph{-x}, it can be difficult to track the cherry-picked record.

Third, to decide whether to merge or to cherry pick, there are suggestions to use instead the flow that merges the changes into the downstream branch and then merge upstream. There are situations that the branches can not be merged as they have to be maintained separately. And because this is a manual process, there is a possibility to merge into the different branches instead of the one intended.  Also, because pull-based software development often involves a number of contributors and those who integrate changes from them must understand all changes and possible conflicts, while cherry-picking is done at the commit level instead of the branch level. It can be done more quickly, and thus more suitable for fixes that required immediate handling.

\section{Tartarian approach}

We present Tartarian, a tool for identifying branches that have a potential to benefit from commits in other branches. Tartarian helps developers to make decisions whether the commit should be cherry picked into a release branch. Tartarian supports Maven and Git by parsing the \emph{pom.xml} configuration file and searching the dependency requirements to construct a \emph{Dependency Graph} that represents the dependencies of the release branches on external libraries. In addition to the original commit message, Tartarian requires developers to provide hashtags to indicate the purpose of commit regarding the dependencies. The \emph{Branch Identifier} matches the dependency graph and the hashtags to identify the applicable branches for the commits. To use Tartarian, a developer can use git \emph{tcommit} operation with some hashtags in the message. For example,

\begin{verbatim}
$git tcommit -m "fixed bug 132215 #bugfix{JDK, 1.8+}"
\end{verbatim}

Tartarian analyses hashtags and uses Branch Identifier to search through the dependency graph to identify the affected releases and makes recommendations to cherry pick. Developers can read the recommendations using,

\begin{verbatim}
$git tcommit -r 
\end{verbatim}

\newpage
\subsection{Dependency Graph}

Tools such as MaX \cite{Srivastava05} can determine control and data dependencies and create a dependency graph to analyse the impact of code changes. Build tools such as Maven, Ant and Gradle maintain dependency configurations and acquire the required libraries during the build. In Maven, the dependencies of a release are maintained in pom.xml. Tartarian parses the pom.xml files and constructs a global dependency graph for all releases present. %For instance, Jetty 9.2.x requires JDK 1.7.0 and Jetty 9.3.x and 9.4.x require JDK 1.8.0. And all releases depend of the javax.servlet-api 3.1.0.

Dependency graph can be defined as a directed graph $G = (V, E)$ where V is a set of modules and E $\subseteq$ V$\times$V is a set of dependencies. $V(X, Y)$ indicates a module X depends on modules Y. Using dependency graph, Tartarian can address the \emph{``dependency hell''} problems such as long chains of dependencies e.g. $V(W, X), V(X, Y), V(Y, Z)$, dependency conflicts e.g. $V(X, Z_1), V(Y, Z_2)$ and circular dependency e.g. $V(X, Y), V(Y, Z), V(Z, X)$.

%where $Z_1$ and $Z_2$ are different versions of a library

\subsection{Tartarian Hashtags}

To communicate through space and time,
developers describe the purpose of a commit in a commit message.
Other developers must read the message to understand the purpose.
If the commit message has insufficient information,
it is difficult for other developers to know what he should do and
whether he should apply the commit into his branch,
despite the fact that the developers may know about the existence of the commit.

We analyse practices and guidelines, such as the JDK migration guide \cite{Oracle:migrate},
which describe the removed and changed APIs, the deprecated list of APIs \cite{Oracle:deprecated}
and the conventional uses of cherry-picking.
We also examine the cherry-picking practices of the Jetty project and
four other open source projects including \emph{cpython, elasticsearch, hadoop, and linux}.
We analyze commit messages that contain the text \emph{``cherry picked from''}
for traceable cherry-picked commits.
The process has two steps. First, for each project, we use R to find term occurrence
in the messages. Second, we use those terms as keywords that appear more frequently
to classify the messages into categories.
Based primarily on the Jetty project, we use qualitative analysis into the detail
of issues and code changes. In table \ref{tartariantag}, we propose an initial set
of hashtags including \emph{bugfix, backport, config, deprecated, improve, inaccessible}
and \emph{removed}, each of which has tag name together with the dependency constraints
such as dependency on external libraries and the versions of those libraries
that the commit is related with.

\scriptsize
\begin{table}[!h]
\caption{Tartarian hashtags}
\begin{center}
\begin{tabular}{|p{1.15cm}|p{2.75cm}|p{2.1cm}|p{.75cm}|}
\hline
Tag & Description & Example & Priority\\
\hline
\#bugfix & A branch will result in an error, without this commit. & \#bugfix\{JDK, +\} & High \\
\hline
\#backport & Backport this commit from a newer release to an older one. & \#backport\{Jetty, 9.2.x\} & Medium \\
\hline
\#config & Changes in configuration which may effect some of the releases. & \#config\{JDK, 1.8+\} & High \\
\hline
\#deprecated & A method is deprecated and there is a better alternative in this commit. & \#deprecated\{JDK, 1.8+\} & Low\\
\hline
\#improve & This commit provides an improvement for a higher quality of code. & \#improve\{Maven, +\} & Low\\
\hline
\#inaccessible & The method is found to be inaccessible, without this commit a branch will not compiled. & \#inaccessible\{JDK, 1.8+\} & High \\
\hline
\#removed & A method is removed but still in backwards compatibility. & \#removed\{JDK, 8+\} & Medium\\
\hline
\end{tabular}
\end{center}
\label{tartariantag}
\end{table}%
\normalsize

These hashtags can be prioritised into three levels of urgency: high, medium and low. A commit with hashtag of high priority must be cherry picked immediately and without them software might fail. A hashtag of medium priority must be dealt with eventually when a developer can. And the hashtag with low priority indicates that the commit may be ignored but to apply it to the release can be beneficial to the quality of software.

\subsection{Branch Identifier}
After a code change is committed to the repository, Tartarian checks if a commit has a relevant hashtag. If so, it analyses the tag and parses for the dependency constraints attached with the tag.
The Branch Identifier represents dependency graph in Neo4j \cite{neo4j}, a database for graph data structure, and translates the constraints into Cypher \cite{cypher}, a query language for graphs, to retrieve the affected release branches. For example, \#bugfix\{JDK, 1.8+\} indicates a commit applicable for the releases that have dependency on JDK 1.8 and later. Each release is then checked for its dependency satisfying the constraints.
With this information, Tartarian can notify developers who are responsible for the affected releases to consider cherry picking this commit into their branches with high priority.

Figure \ref{fig_tartarian} depicts how a commit in Jetty 9.4.x with a hashtag \emph{\#bugfix\{JDK,1.8+\}} is resolved with Jetty 9.2.x and 9.3.x where release 9.2.x has dependency on JDK1.7 and release 9.3.x on JDK1.8. The commit is resolved with release 9.3.x and a developer maintaining this release is notified with this commit having a high priority.

\begin{center}
\includegraphics[width=3.25in]{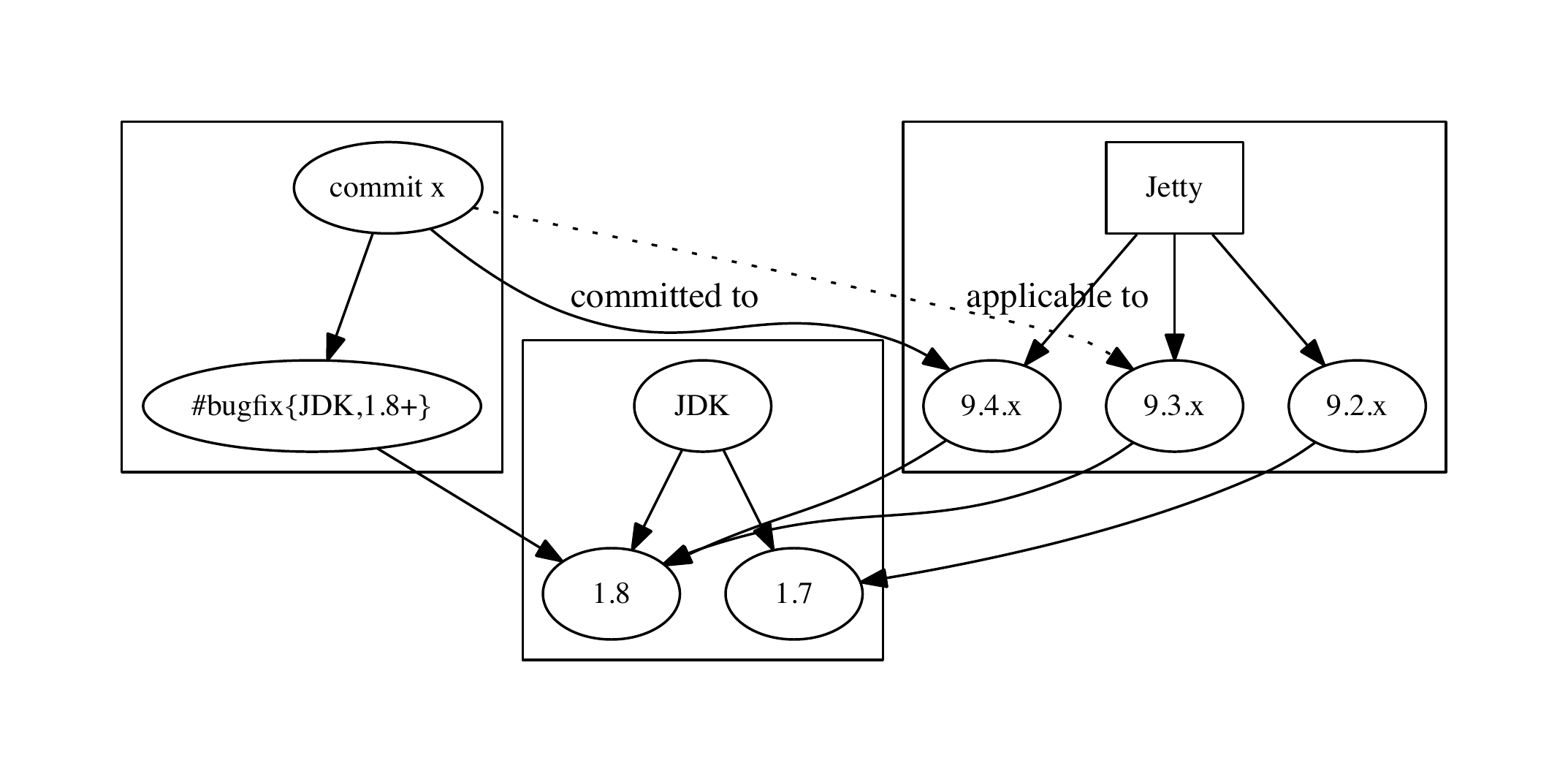}
\captionof{figure}{How Tartarian works}
\label{fig_tartarian}
\end{center}

%\newpage
\section{Case Studies}

This section illustrates the cherry-picked commits in Jetty, a web server and servlet container. Since 2016, Jetty repository is maintained at Github under \url{https://github.com/eclipse/jetty.project}. We use data from \emph{jetty-9.2.x, jetty-9.3.x} and \emph{jetty-9.4.x} for this study.

To analyse cherry-picked commits in the Jetty project, we identify the commit messages that contains the phrase \emph{``cherry picked from \textless commit id\textgreater''}. We search with the term \emph{``cherry picked''} and \emph{``cherry''} to trace all cherry-picked commits using the SourceTree tool. When we found such commit, we explore the intention of the commit by looking at the issue that tends to cause this commit, the full commit message and the \emph{diff} of source code of both the original commit and the cherry-picked one.

We use ``git branch --contains \textless commit id\textgreater'' to identify all branches that contain the commit and verify that the commit is indeed cherry picked. Once we have the \emph{branch id}, we check the commit and source code in Github to analyse the changes.

In the Jetty repository 9.2.x, 9.3.x and 9.4.x, SourceTree returns 19 messages from the search with keyword \emph{``cherry picked''} i.e. the cherry picked commits and 27 messages with keyword \emph{``cherry''} that indicates other commits such as \emph{``cherry pick correction''}, \emph{``undo cherry pick''}, \emph{``cherry pick cleanup''} etc.

\subsection{\#deprecation of older methods}

It is possible that the newer version of external libraries might provide a better alternative to code. For example, in issue 1135 ``Avoid allocations from Method.getParameterTypes() if possible \#1135'' found at \url{https://github.com/eclipse/jetty.project/issues/1135}, the contributor suggests to change all the method calls as below.

\begin{verbatim}
       -if (m.getParameterTypes().length != 0)
       +if (m.getParameterCount() != 0)
\end{verbatim}

There was a pull request. It was suggested that there should have been a merge into release 9.3.x but the commit was pulled instead into \emph{master}. Then, it was cherry picked into release 9.3.x and then merged into release 9.4.x. %This is due to the fact that pull-based workflow can be different from team to team.

With Tartarian, the cherry-picking can be semi-automated and thus better supported. The commit can be tagged with ``\#deprecated\{JDK, 1.8+\}'', suggesting that this change will affect the releases that depend on JDK 1.8 onwards, without the knowledge about other branches. Tartarian can search from the dependency graph for the affected release branches i.e. Jetty 9.3.x and 9.4.x in this case, both having dependencies on JDK 1.8, and then notify the developers to consider cherry picking this commit without the developers having to communicate.

\subsection{\#bugfix}

In commit \emph{fa53b11} in release 9.4.x, the commit message is \emph{``fixed reset of DispatcherType''}. The original source code has a \emph{finally} block at the end of the \emph{case} selection which resets the \emph{DispatcherType} to \emph{null} for every possible case. This error was removed and each case \emph{has} its own \emph{finally} block instead. The commit was cherry picked to release 9.3.x as commit \emph{56afc2}, tagged as \emph{``481554 DispatcherType reset race.''}

Because the bug fix concerns programming logic, it is likely that this change should apply to all release branches. However, this commit is not cherry picked into release 9.2.x.

Tartarian can help to facilitate this change. Developers can commit the change with a tag \emph{``\#bugfix\{JDK, +\}''} to specify that the commit may be applied to all release branches having dependency on any version of the JDK because \emph{finally} is introduced into Java programming language since the early version of the JDK. The tag would be checked and recognised as \emph{high} priority. The developers taking care of all other release branches then would be notified with this change and can make decision to include the change into their branches.

\subsection{\#configuration changes}
%https://github.com/eclipse/jetty.project/commit/521cc6

The commit \emph{521cc6} containing the message \emph{``Some javadoc plugin configuration updates (cherry picked from commit 240c217)''} describes the changes in software configuration that may affect various release branches that depend on the external libraries specified in the configuration. In the original commit (240c217), the configuration was removed from \emph{jetty.project/pom.xml} specifying that the remove was required for JDK 8u121 or later. And the cherry-picked commit (i.e. commit 521cc6) is modified to support the release branch with a dependency on JDK1.8 onward. The two commits suggest that implicit knowledge about the JDK version that affect the configuration of Jetty are specified traditionally in the comment.

With Tartarian, the commit message can be tagged with \emph{\#config\{JDK,1.8+\}} instead. By doing so, other release branches with dependency on JDK1.8 can be advertised with the updates in the configuration and developers can make a decision to cherry pick the commit into their branches.

\subsection{ \#backport from a newer release}

Backport is essential for backward compatibility and for the code in newer branch to produce the same output as those in older branches. In commit \emph{5c0906e} in release 9.3.x, the commit message is \emph{``474454 - Backport permessage-deflate from Jetty 9.3.x to 9.2.x + post cherry-pick merge cleanup''}. The commit message does not contain the hash of the original commit, without this information, it is very difficult to identify the original commit.

To backport from a branch to an older release branch, Tartarian uses \emph{\#backport\{Jetty, 9.2.x\}} to indicate that a commit can be used to backport into the 9.2.x release branch. Developers taking care of the release branch then would be notified with this change and can make decision to backport this change into their branch in a more systematic way.

%Tracing this change, we found that the permessage-deflate issue originates since commit \emph{e0e00b0} with the commit message \emph{``430418 - Jetty 9.1.3 and Chrome 33 permessage-deflate do not work together + Updating Compress/PerMessageDeflate extensions for latest spec document, http://tools.ietf.org/html/draft-ietf-hybi-permessage-compression-18 Intra-frame tail 0000FFFF is now being preserved for permessage-deflate''}

\subsection{\#improve the quality of software}
%\url{https://github.com/eclipse/jetty.project/commit/be5bb05}

This case illustrates the improvement to code. Commit \emph{be5bb05} with message \emph{``Disabling javadoc, deploy, findbugs in /tests/ (cherry picked from commit ad1512d)''} proposes changes to the \emph{tests/pom.xml} file so that the maven-deploy-plugin and maven-javadoc-plugin are skipped and that the projects are not deployed during test and that no java doc is generated, thus the overall performance can be improved. The pom.xml file is changed in release 9.4.x and is cherry picked into 9.3.x to disable the unnecessary plugins in the \emph{org.eclipse.jetty.tests} package. In release 9.2.x there is no configuration for the two plugins so it is unnecessary to apply the change.

Using Tartarian, the tag for quality improvement can be written as \emph{\#improve\{Maven, +\}} to notify other release branches that depend on Maven for potential code improvement. However, in this particular case, for some reasons, the configuration in release 9.2.x does not contain the plugins of interest. Therefore the developers responsible for the release branch may choose to ignore the cherry-pick if they consider this as unnecessary or they may choose to apply the commit to make the file consistent among all branches.

The \emph{\#inaccessible} tag can be used in the situations similar to those of the \emph{\#bugfix} tag, as the code will not be compiled or executed without these inaccessible methods; therefore, equivalent to a bug in the code. And the \emph{\#removed} tag can be used in the same situations as those of the \emph{\#deprecated} tag, as the code can still be compiled and executed but may need to be updated at some point in the future.

\section{Cherry-picking in other software projects}

%git log --pretty=oneline --branches --grep "cherry picked from"
%We choose the projects based on two criteria. Firstly, the projects must maintain several release branches. And secondly, there must be a sufficient number of cherry-picking commits.
We found that many software projects have long-running release branches to maintain multiple versions of the software. Examples include python 2.7, 3.5 and 3.6. Due to a large numbers of external libraries only support Python 2.x when the language itself has progressed to Python 3.x, the Python community consequently has to maintain both versions of the language for several years. Elasticsearch and Hadoop maintain a number of release branches. For Linux, while the project maintain only one master development branch on Github, there are 513 releases as of June 2017.

As discussed in section 5.2, we try to generalise the cases found in the Jetty project to other open source projects to examine the pattern of cherry-picking. Based on this high-level analysis, the cherry-picked commits can be categorized into three groups: \emph{``bug fixes''} for commits related to bugs, \emph{``backports''} for commits regarding backward compatibility, and \emph{``code maintenance''} for the messages that contain verb words such as \emph{`add', `remove', `improve', `change' etc.} At this stage, without analysing the source code in a detail, it is difficult to further classify the commits in the \emph{code maintenance} into the subgroups at the level of our defined tags in the case study. We plan to further develop this in the future work. %Table \ref{projects} shows the classification of cherry-picked commits of some open source projects. Note that the three groups can be overlapped and some commits are not in the three groups.

\scriptsize
\begin{table}[h!]
\caption{Cherry-picks in four major open source projects}
\begin{center}
\begin{tabular}{|l|l|l|l|l|l|}
\hline
Project 	& release		& total 			& bug  	& back- 	& code 	\\
 		& branches 	& 		& fixes 	& ports 	& maintenance \\
\hline
cpython 				& 2.7, 3.5, 3.6	& 461	& 149 	& 10 		& 264 	\\
(python/cpython)		&			&		& (32.32\%) 				& (2.17\%)				& (57.27\%)	\\
\hline
elasticsearch 			& 2.0-2.4, 		& 337	& 49		& 1 			& 269 	\\
(elastic/elasticsearch)	& 5.0-5.4, 5.x	&		& (14.54\%)		& (0.30\%)		& (79.82\%)		\\
\hline
hadoop 				& 2, 2.6-2.8.1,	& 7150	& 1431 	& - 		& 2095	\\
(apache/hadoop) 		& 3.0.0-alpha1,	& 	 	& (20.01\%)		&  		& (29.30\%) 		\\
			 		& 2 and 3 	& 	& 		&  		& 		\\
\hline
linux 					& -			& 534 	& 98		& - 		& 119 	\\
(torvalds/linux)			&			&		& (18.35\%)			&		& (25.81\%)		\\
\hline
\end{tabular}
\end{center}
\label{projects}
\end{table}%
\normalsize
\section{Conclusion and Future Work}

We propose that, together with commit messages, developers may use hashtags as metadata to help specifying the intention of a commit. These hashtags, when used with the Tartarian tool, can identify the release branches that a commit may be reused. Doing so can help developers to efficiently maintain software without the need to acquire implicit knowledge of the release branches. 

Despite the fact that dependencies on JDK are used in most of the case studies in this paper, we believe that Tartarian can be applied to dependency constraints in general, which we will explore in our future work. In addition, it can be argued that the satisfaction of dependency constraints alone is not sufficient for the cherry-picked commit to fit the branch. Cherry-picking requires semantic checking to establish that the cherry-picked commit will not cause unintentional faults or build breaks in the branch. A feature that checks for these semantic constraints is needed in Tartarian.

\section{Acknowledgement}
The authors are grateful to the reviewers for their helpful suggestions and would like to thank the University of the Thai Chamber of Commerce for supporting this project.

\balance
\bibliographystyle{ACM-Reference-Format}
\bibliography{sigproc}

\end{document}